\documentclass[pdflatex,sn-mathphys-num]{sn-jnl}% Math and Physical Sciences Numbered Reference Style 
\usepackage{graphicx}%
\usepackage{multirow}%
\usepackage{amsmath,amssymb,amsfonts}%
\usepackage{amsthm}%
\usepackage{mathrsfs}%
\usepackage[title]{appendix}%
\usepackage{xcolor}%
\usepackage{textcomp}%
\usepackage{manyfoot}%
\usepackage{booktabs}%
\usepackage{algorithm}%
\usepackage{algorithmicx}%
\usepackage{algpseudocode}%
\usepackage{listings}%
\usepackage{array}
\usepackage{xr}
\usepackage{geometry} 
\newcolumntype{P}[1]{>{\centering\arraybackslash}p{#1}}
\newcolumntype{M}[1]{>{\centering\arraybackslash}m{#1}}

\newcommand*{\addFileDependency}[1]{% argument=file name and extension
\typeout{(#1)}% latexmk will find this if $recorder=0
\IfFileExists{#1}{}{\typeout{No file #1.}}
}\makeatother

\newcommand*{\myexternaldocument}[1]{%
\externaldocument{#1}%
\addFileDependency{#1.tex}%
\addFileDependency{#1.aux}%
}
\myexternaldocument{SI}

\theoremstyle{thmstyleone}%
\begin{document}

\title[Article Title]{Impacts of EPA Power Plant Emissions Regulations on the US Electricity Sector}

\author[1]{\fnm{Qian} \sur{Luo}}\email{ql7299@princeton.edu}
\author*[1]{\fnm{Jesse} \sur{Jenkins}}\email{jdj2@princeton.edu}

\affil[1]{\orgdiv{Andlinger Center for Energy and the Environment}, \orgname{Princeton University}, \orgaddress{\city{Princeton}, \postcode{08540}, \state{NJ}, \country{US}}}

%(250 words)}
\abstract{

Taking aim at one of the largest greenhouse gas emitting sectors, the US Environmental Protection Agency (EPA) finalized new regulations on power plant greenhouse gas emissions in May 2024. These rules take the form of different emissions performance standards for different classes of power plant technologies, creating a complex set of regulations that make it difficult to understand their consequential impacts on power system capacity, operations, and emissions without dedicated and sophisticated modeling. Here, we enhance a state-of-the-art power system capacity expansion model by incorporating new detailed operational constraints tailored to different technologies to represent the EPA's rules. Our results show that adopting these new regulations could reduce US power sector emissions in 2040 to 51\% below the 2022 level (vs 26\% without the rules). Regulations on coal-fired power plants drive the largest share of reductions. Regulations on new gas turbines incrementally reduce emissions but lower overall efficiency of the gas fleet, increasing the average cost of carbon mitigation. Therefore, we explore several alternative emission mitigation strategies. By comparing these alternatives with regulations finalized by EPA, we highlight the importance of accelerating the retirement of inefficient fossil fuel-fired generators and applying consistent and strict emissions regulations to all gas generators, regardless of their vintage, to cost-effectively achieve deep decarbonization and avoid biasing investment decisions towards less efficient generators.}


\maketitle

\section{Introduction}\label{sec1}

\begin{figure}[h]
    \centering
    \includegraphics[width=1\textwidth]{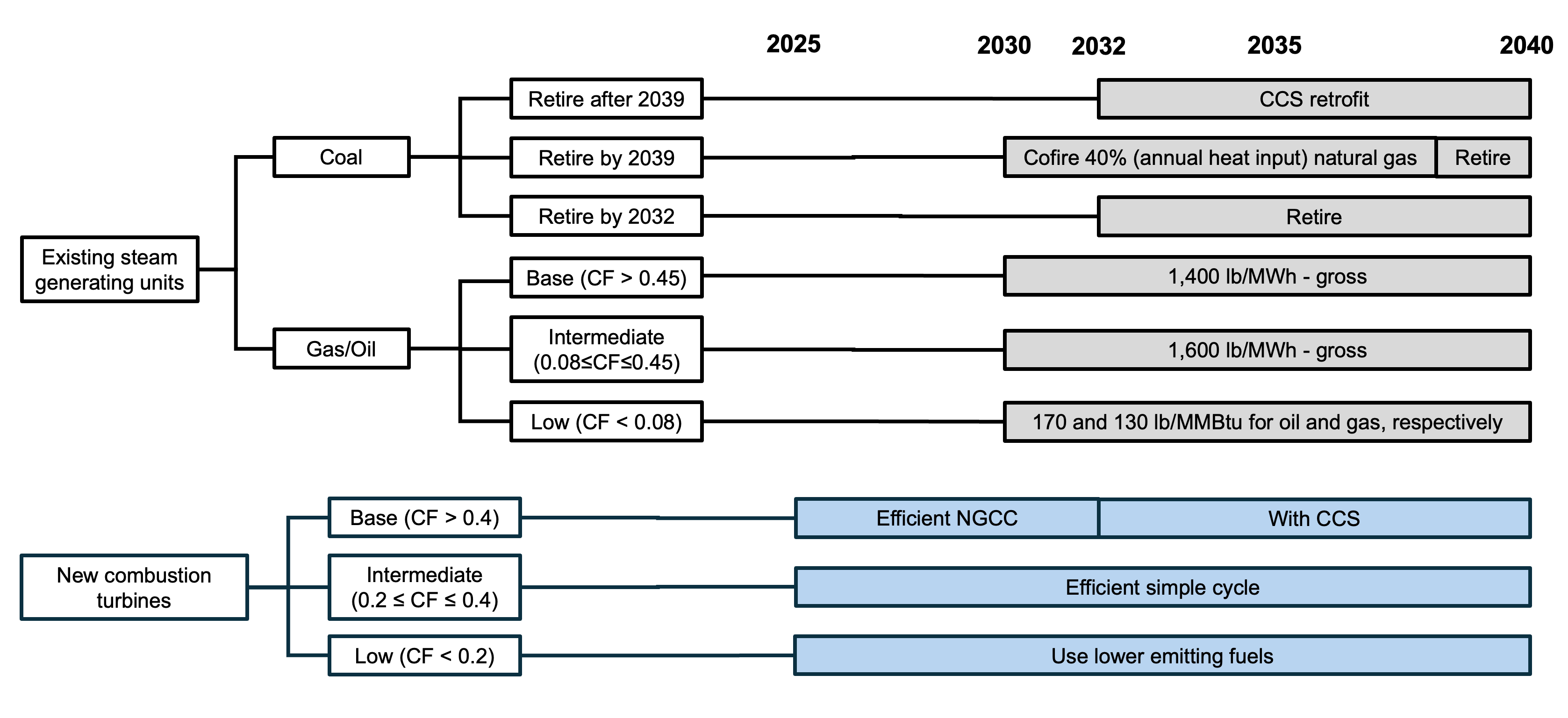}
    \caption{CO\textsubscript{2} emission mitigation options in each modeling period for different generator types in GenX.}
    \label{fig:EPA_rules}
\end{figure}

To keep global warming ``well below'' $2^{\circ}$C (compared to the pre-industrial levels), nearly 200 nations agreed to cut greenhouse gas (GHG) emissions under the Paris Agreement, but no specific guidance on how to effectively reduce emissions was provided\cite{paris}. In the United States (US), the Inflation Reduction Act (IRA) of 2022 and Infrastructure Investment and Jobs Act of 2021 collectively deploy over \$500 billion of tax credits, grants, rebates, and loan guarantees to incentivize clean energy investment and reduce greenhouse gas emissions in the US\cite{ira}. These policies are likely to reduce US greenhouse gas emissions to 33 - 40\% below 2005 levels by 2030 and 43 - 48\% by 2035 \cite{bistline2023emissions}, falling short of the United States' nationally determined contribution (NDC) under the Paris Agreement (50-52\% below 2005 emissions in 2030)\cite{ndc_short} and long-term strategy (net-zero by 2050)\cite{ndc_long}. To further reduce GHG emissions, the Biden Administration has proposed and finalized several additional sectoral emissions regulations (including transportation, oil and gas methane emissions) and efficiency standards. The regulations on CO\textsubscript{2} emissions for fossil fuel-fired electricity generating units (EGUs) recently finalized by the US Environmental Protection Agency (EPA) under Section 111 of the Clean Air Act are among the most significant of these regulations\cite{epa_final}. The EPA power plant regulations introduce different emissions performance standards for several different classes of generators with different deadlines to meet those requirements. For example, existing coal-fired steam generators must meet emissions rates equivalent to either equipping carbon capture and storage (CCS) by 2032 or co-firing natural gas by 2030, depending on their planned retirement date, while new natural gas (NG) combustion turbines have to equip CCS (or meet an equivalent emissions rate) by 2032. Operational utilization rates (capacity factors, CF) also affect how these regulation are applied to specific generators. For example, new gas generators that run infrequently in a year (``non-baseload generators'', defined as generators with capacity factor below 40\%) would not be subject to this CCS requirement. Figure \ref{fig:EPA_rules} summarizes this complex set of requirements.

EPA proposed its first CO\textsubscript{2} emission regulations for existing fossil fuel-fired power plants in 2014, known as the Clean Power Plan (CPP)\cite{cpp}. CPP set state-specific emission reduction targets and provided flexibility for states to determine how to reach these goals (e.g., with mass- or rate-based emissions limits and the option for emissions permit trading across linked states). Although many studies show that CPP would have resulted in significant climate and public health benefits \cite{cardell2015targeting,murray2015regulating, driscoll2015us,davis2016state}, the proposed rule was curtailed by the Supreme Court due to ``the lack of the authority'' to set a sector-wide GHG emissions performance standard based on generation shifting\cite{epa_wv}. 

Unlike CPP, new GHG regulations finalized by EPA in 2024 provide different generator-level (``inside-the-fenceline'') performance standards for power plants with different technologies, vintages, and utilization rates (capacity factor). The complexity of the proposed EPA rules makes it difficult to estimate how the new regulations would affect the US power system, including capacity investment, retrofit and retirement decisions, operations, and greenhouse gas emissions. In this work, we extend a state-of-the-art, open-source power system capacity expansion model, GenX\cite{jenkins2017enhanced,genx}, by incorporating detailed operational constraints tailored to different technologies to represent the EPA rules, with an aim to answer the following four research questions:

    1. What are the effects of the rules targeting each class of generators and how do they interact with each other in various combinations?

    2. What are the emission and economic impacts driven by the key changes from proposed to final rules (e.g., a decision to delay regulations on existing gas generators)?
    
    3. How sensitive are expected emissions outcomes to key uncertainties related to future fuel costs, renewable resource availability, and tax credits?
    
    4. Are there alternative regulatory strategies that can be applied to power plant emissions that could yield additional emissions reductions or improve the average cost of emissions mitigation?

To answer these questions, we formulate new constraints to reflect the various compliance pathways and emissions performance standards for each class of generators and model a range of scenarios with different combinations of rules under different uncertainties. This work represents the first independent research on the effect of the final EPA power plant rules applying a detailed electricity system planning model. Our modeling results show that the majority of emissions reductions are associated with the regulation on coal-fired generators. Constraints on new natural gas generators help reduce emission further but lower the overall system efficiency, reducing emissions at a very high incremental cost per ton of CO$_2$ avoided. Additionally, we determine that applying more consistent emissions regulations to all gas generators, regardless of their vintage, would avoid biasing investment decisions towards less efficient generators and improve the economic efficiency of the proposed rules.

\section{Results}\label{sec2}
\subsection{EPA power plant emissions rules reduce US power sector GHG emissions by 35\% in 2040 but only some of the rules play a major role.}

\begin{figure}[h]
    \centering
    \includegraphics[width=1.1\textwidth]{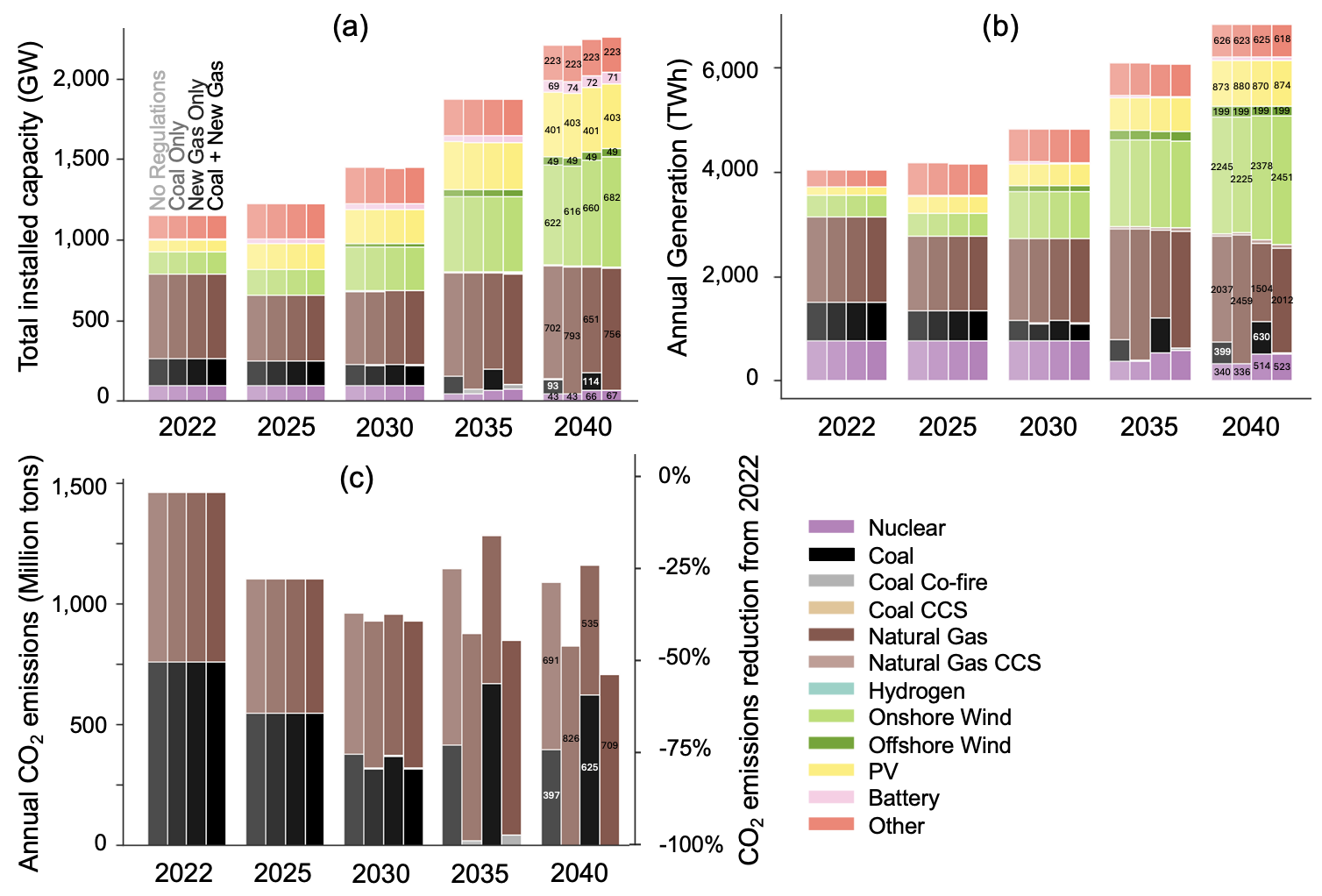}
    \caption{Total installed capacity (a), electricity generation (b), and CO\textsubscript{2} emissions (c) by technology in each period. 2022 data is from EIA and other periods are modeling outputs.}
    \label{fig:capGenEmis_all}
\end{figure}

The EPA finalizes two separate rules targeted at two groups of fossil fuel-fired generators: 1) existing steam generating units (which we refer to as the ``Coal" rule because most of the existing steam generating units are coal fired); and 2) new combustion turbines, including combined cycles (``New Gas"). To understand the impacts of each rule on the system, we design three scenarios: ``Coal Only", ``New Gas Only", and ``Coal + New Gas". We also include a benchmark case (``No Regulations''), where no EPA rules are considered. All cases include relevant tax credits enacted by IRA and projected electricity demand accounting for IRA impacts on electrification of transportation and heating\cite{repeat}. Figure \ref{fig:capGenEmis_all} shows installed capacity, electricity generation, and CO\textsubscript{2} emissions across all technology types under the four scenarios. Figures \ref{fig:CapExisting} and \ref{fig:tx_and_demand} reflect the existing power system used in this study and Figure \ref{fig:2021} benchmarks model performance relative to historical outcomes.

Imposition of rules on existing coal-fired generators has the most significant impact on installed capacity, generation, and emissions outcomes. This rule requires coal-fired generators with no planned retirement before 2039 to be retrofitted with CCS from 2032 and generators that plan to retire before 2039 to co-fire at least 40\% natural gas (on a heat input basis) from 2030, or meet equivalent emissions rates (though we assume herein that no better compliance options are available). No additional constraints will be applied to coal-fired generators with scheduled retirement before 2032. Constraints on coal generators result in 28 GW of coal retrofit to co-fire with natural gas in 2035 and additional 90 GW natural gas turbines capacity in 2040, but almost no coal-fired generators choose to add CCS equipment even with the tax credits that IRA provides for CCS (``45Q"; Section \ref{sec:sensitivity} presents a more detailed discussion about the impacts of how 45Q is modeled in this study). Regulations on coal-fired generators significantly reduce installed coal capacity and associated generation, accounting for the majority of emission reductions from the EPA rules. Compared with the ``No Regulations'' case, ``Coal Only'' alone reduces CO\textsubscript{2} emissions from 2035 onwards (during which the EPA regulations for coal-fired generators apply) by 24\% in each period. Emissions in 2035 and 2040 are 270 and 260 million tons of CO$_2$ (MtCO$_2$) lower, respectively, than the ``No Regulations'' case, reaching 40\% and 43\% below 2022 power sector emissions. Regulations on existing oil/gas steam turbines do not have significant impacts on system emissions as most of them retire or run at very low utilization levels under the ``No Regulations" scenario due to low efficiency and high marginal costs of generation.

\begin{figure}[h]
    \centering
    \includegraphics[width=1.1\textwidth]{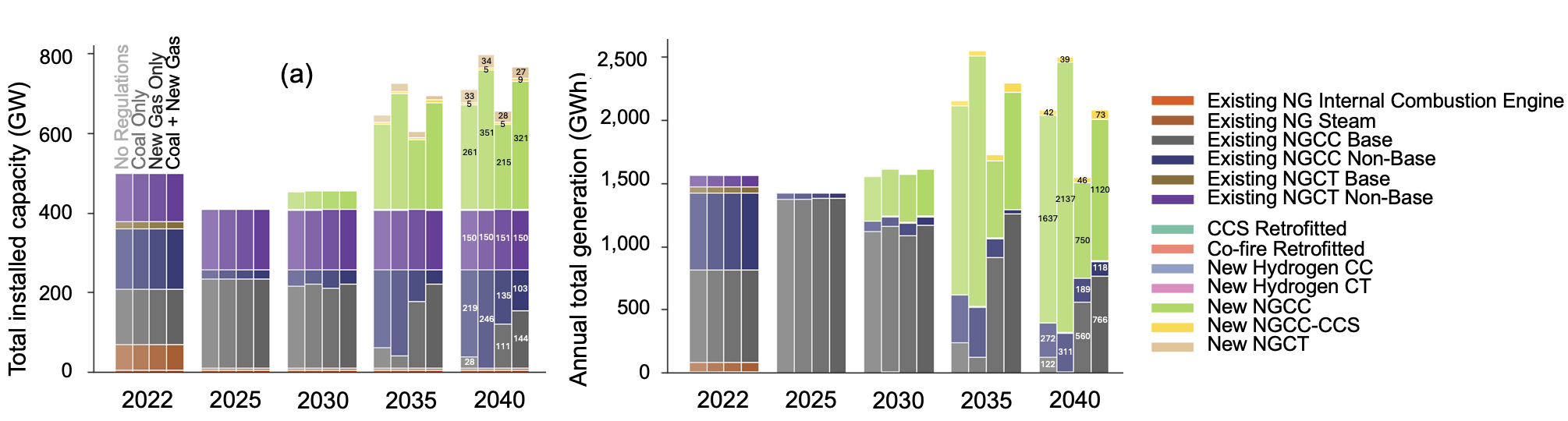}
    \caption{Total installed capacity (a) and electricity generation (b) for natural gas and H\textsubscript{2}-fueled generators in each period. 2022 data is from EIA and other periods are modeling outputs. (Base: capacity factor $>$40\%)}
    \label{fig:capGen_NG}
\end{figure}

When only adding regulations on new combustion turbines without the coal regulations (``New Gas Only"), new NG turbines must either keep capacity factor below 40\% or install CCS in 2032. The imposition of these constraints on new gas turbines effectively increases the cost of new gas capacity, resulting in more generation from coal, existing nuclear, and renewable resources. Although emissions from the combustion of natural gas are reduced by 23\% in 2040, increased use of coal makes the ``New Gas Only'' scenario exhibit even greater total emissions than the ``No Regulations'' benchmark. When this gas rule is added to ``Coal Only'' (``Coal + New Gas''), as in the finalized EPA rules, it incrementally reduces CO\textsubscript{2} emissions by 115 MtCO$_2$ in 2040 (reaching 51\% below 2022 emissions). In this case, 66 GW more onshore wind is added to the system on the basis of ``Coal Only'', with natural gas capacity reduced by 37 GW. Additionally, we observe 24 GW less nuclear retirement compared with the ``Coal Only'' scenario. However, there are still substantial emissions from the combustion of natural gas under this scenario. Therefore, we break down NG-fired generators into several subgroups based on their technology and operational status to understand how regulations on new NG turbines impact emissions from the combustion of natural gas (Figure \ref{fig:capGen_NG}). 

Under the final EPA rules (``Coal + New Gas''), more new gas capacity is added through 2040 than the benchmark ``No Regulations'' case but produces less generation. New installed capacity of gas combined-cycle turbines without CCS (``New NGCC'') increases by 23\% and generation from these units is reduced by 32\%. Although most gas capacity is still New NGCC (321 GW, 42\% of total installed gas combustion capacity in 2040), these units operate below 40\% capacity factor to avoid emissions standards for new baseload gas plants requiring CCS and provide 54\% of electricity generation from natural gas combustion in 2040. In comparison, due to higher efficiency, New NGCC provides 86\% of all gas-fired generation under ``Coal Only'' with 30 GW more installed capacity of New NGCC than ``Coal + New Gas.'' Meanwhile, electricity generation from existing gas turbines almost doubles after Rule ``New Gas'' is included. A small amount of new NGCC with CCS (New NGCC-CCS) is added under ``No Regulations'' (5 GW) and the regulation on new gas turbines increases this to 9 GW. However, the overall shift in generation towards less efficient existing gas-fired generators caused by the new gas regulations increases emissions from the combustion of natural gas even compared with the ``No Regulations" scenario (Figure \ref{fig:HR_NG}). 

\subsection{Compared with proposed rules, final EPA power plant rules result in slightly more emissions in 2040 but a more efficient system and lower average cost of mitigation.}

EPA proposed power plant GHG rules in May 2023\cite{epa_propose} and finalized the rules in May 2024\cite{epa_final}. There are a few major differences between the proposed and final versions: 1) the finalization of rules on existing combustion turbines was delayed as EPA plans to take ``a new, comprehensive approach to cover the entire fleet of natural gas-fired turbines''\cite{epa_delay}; 2) the definition of baseload NG-fired generators was modified from a capacity factor $>$50\% to $>$40\% ; 3) a hydrogen (H\textsubscript{2}) co-firing pathway in the best system of emissions reduction (BSER) was removed for new non-peak gas turbines; 4) a subcategory of existing coal steam turbines (coal-fired generators that plan to retire by 2035, which must run with an annual utilization rate below 20\% under proposed rules) was removed; and 5) the compliance date of coal CCS retrofit was delayed from 2030 to 2032. To evaluate the impacts of these changes from proposed to final rules, we compare impacts on capacity, operations, and emissions under the finalized rules (``Coal + New Gas''), a possible scenario including regulations on existing gas plants (``Coal + All Gas''),  and the original proposed rules (``Proposed Rules,'' see details in Figure \ref{fig:EPA_rules_proposal}). Under ``Coal + All Gas,'' we extend the finalized rules requiring CCS for new baseload gas turbines to large, baseload existing gas turbines (capacity $>$300 MW and annual capacity factor $>$40\%), similar to the proposed rules.

Although regulations on existing gas generators have almost no impact on installed capacity of either gas or other resources under ``Coal + All Gas'', total generation from gas turbines decreases as some large existing NGCC generators need to reduce utilization levels below 40\% to avoid CCS retrofit from 2032 onwards (Figure \ref{fig:proposal}). However, this only reduces overall emissions by 16 and 22 MtCO$_2$ in 2035 and 2040, respectively, compared with ``Coal + New Gas''. In 2035, the existing gas turbine capacity is 427 GW in our modeling, only 23 GW of which would be regulated (e.g., with capacity size $>$300 MW and capacity factor $>$40\%). Therefore, we find that EPA's decision to delay finalization of regulations on existing NG turbines is likely to have minimal negative impacts on emissions. We explore several other possible implementations of existing gas rules in Section \ref{sec:alternatives} below.

\begin{figure}[h]
    \centering
    \includegraphics[width=1.0\textwidth]{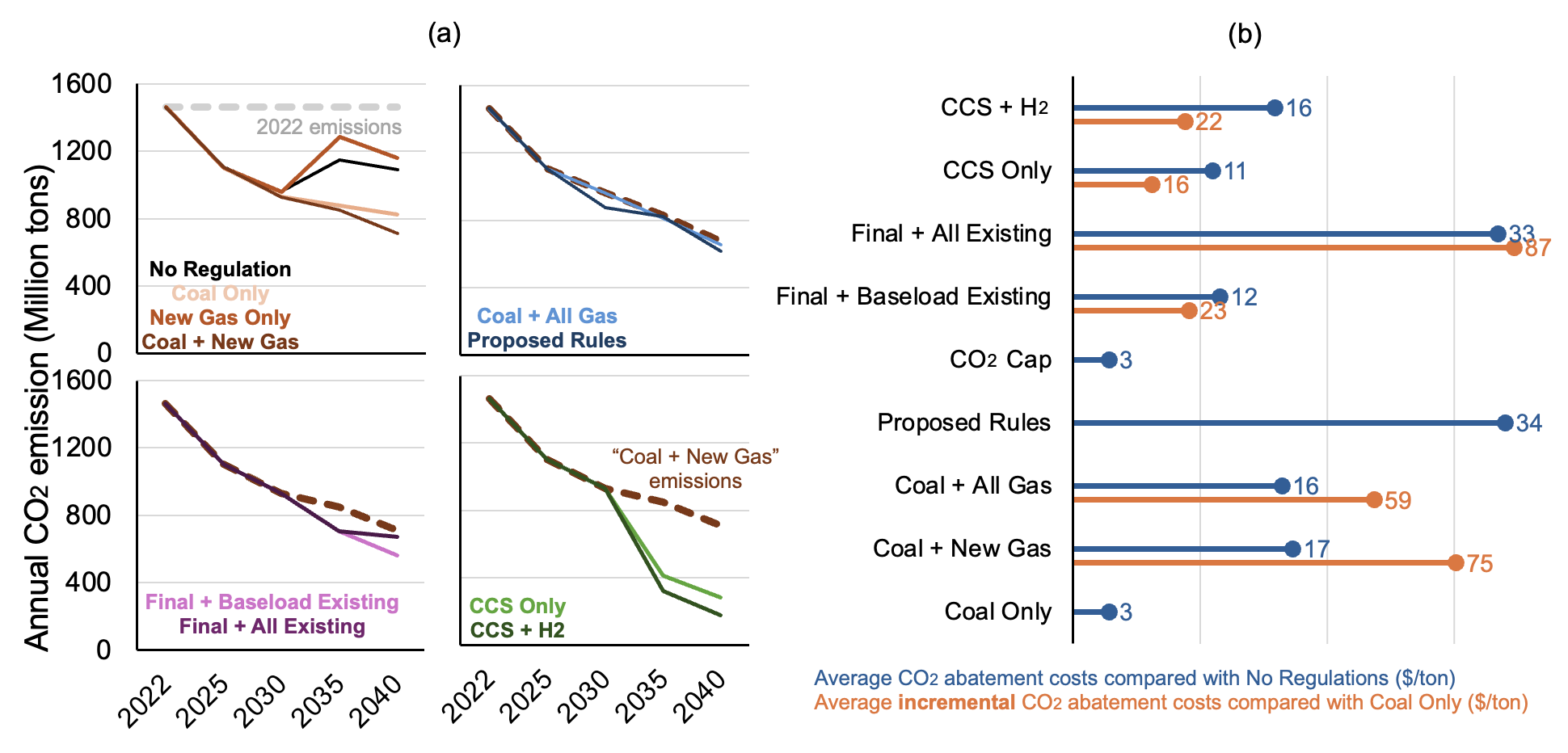}
    \caption{Annual CO\textsubscript{2} emissions (a) and average abatement costs (b) under all scenarios. Abatement costs of ``New Gas Only'' are not listed because it has emission increase. Detailed descriptions of alternative strategy scenarios (``Final + Baseload Existing", ``Final + All Existing", ``CCS Only", and ``CCS + H\textsubscript{2}") are included in Table \ref{tab:caseDescr}.}
    \label{fig:emisByCase}
\end{figure}

Compared with the proposed rules, the most significant difference is driven by the delayed compliance date of coal CCS retrofit (to 2032) and the removal of the sub-category of coal generators that plan to retire by 2035 (required to operate $<$20\% capacity factor). As a result, finalized rules have significantly more generation and emissions from coal in 2030 (Figure \ref{fig:proposal}). In later periods (2035 and 2040), without H\textsubscript{2} co-firing requirements for non-peaker gas turbines, final rules have more natural gas and less renewable capacity, and thus about 40 million tons more emissions in 2040, even after relaxing the standard of baseload generators to cover more NG capacity. However, finalized rules will lead to a more efficient system in terms of gas turbine operations due to less generation from inefficient existing gas generators (Figure \ref{fig:HR_NG}). Therefore, final rules have a lower emission abatement cost compared with the proposed rules (Figure \ref{fig:emisByCase}).

\subsection{EPA rules consistently reduce GHG emissions across sensitivities for renewable resource availability, fuel prices, and tax credit treatment.}\label{sec:sensitivity}
\begin{figure}[h]
    \centering
    \includegraphics[width=1.0\textwidth]{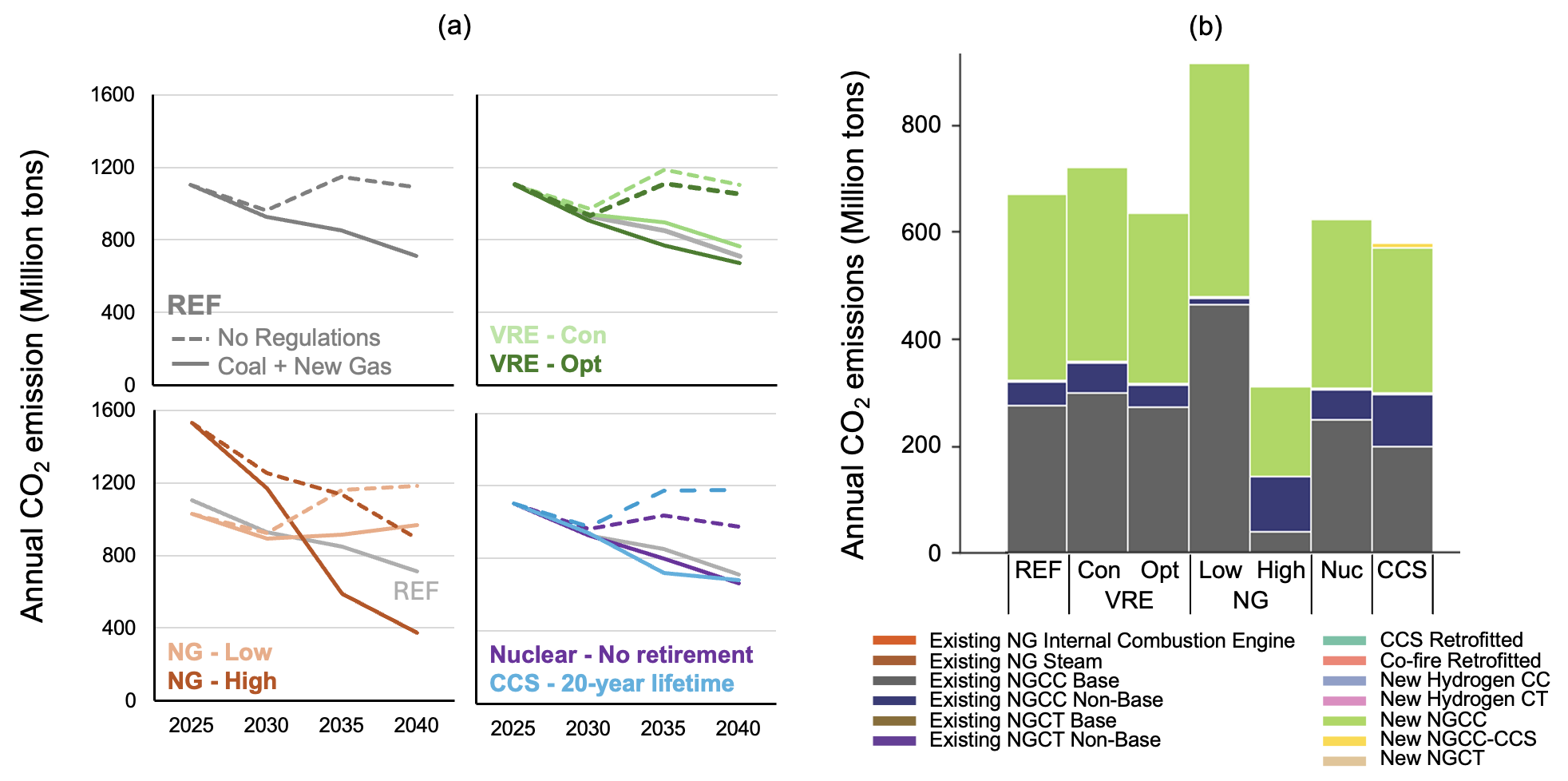}
    \caption{Annual CO\textsubscript{2} emissions in each period (a) and emissions released from different types of natural gas or H\textsubscript{2}-fired generators in 2040 under the ``Coal + New Gas'' rules under different sensitivity scenarios. (Base: capacity factor $>$40\%)}
    \label{fig:emis_sensitivity}
\end{figure}

To assess how future uncertainties might impact the effectiveness of these rules, we evaluate the final rules (the ``Coal + New Gas'' scenario) under three groups of sensitivity analyses, exploring the impacts from availability of renewable resources, prices of fossil fuels, and the treatment of tax credits provided by IRA (Figures \ref{fig:emis_sensitivity} and \ref{fig:capGenEmis_sensitivity}).

In addition to the assumptions used by the base scenario (``REF''), we include another two wind and solar growth limit scenarios: optimistic (``VRE-Opt'') and conservative (``VRE-Con'') (Table \ref{tab:vre} and Figure \ref{fig:VREgrowth}). As building renewable capacity in some regions can be expensive due to transmission expansion investments, the assumed maximum rate of growth in wind and solar capacity additions is likely to have a limited impact on installed capacity and electricity generation of gas turbines. However, we find that fuel prices have strong impacts on system emissions. We obtain low, reference, and high natural gas prices—\$2.8/MMBtu, \$3.9/MMBtu, and \$5.9/MMBtu (national average in 2040), respectively—from the U.S. Energy Information Administration's \textit{Annual Energy Outlook 2023}\cite{aeo2023} (Figure \ref{fig:fuel_prices}). Interestingly, lower NG prices lead to lower emissions in early periods (2025 and 2030) as they result in less generation from coal (and more from gas) while higher NG prices have lower emissions in later periods (2035 and 2040). Overall, compared with ``REF'' under the final rule scenario, lower NG prices result in 36\% more emissions in 2040 (34\% below 2022 emission levels) and higher NG prices result in 47\% fewer emissions in 2040 (74\% below 2022 emission levels).

We also consider uncertainties related to tax credits provided by IRA, including the extension of ``45U'' production tax credit for existing nuclear power plants and the treatment of ``45Q'' tax credit for carbon sequestration. We assume in our ``REF'' cases that existing nuclear units will not retire prior to 2032 due to 45U, but permit the model to choose economic retirements in the 2035 and 2040 planning stages, as the current policy is scheduled to expire after 2032. It is plausible that some form of policy support may be extended to avoid nuclear retirements, so in order to understand the impacts of the interactions between nuclear retirements and other energy resources, we repeat all the analyses above without allowing any nuclear retirement (``Nuclear - No retirement''). With renewable growth limits and fuel prices consistent with the ``REF" scenario, we find that preventing any nuclear retirement after 2032 would reduce CO\textsubscript{2} emission by 48 MtCO$_2$ in 2040 (reaching 55\% below 2022 emissions level). Additionally, as we optimize capacity expansion in GenX in a myopic way (see Methods), it is necessary to annuitize capital expenditures. Thus, to ensure equivalent treatment of production subsidies, we model the 45Q CCS credits, which are available at \$85/tCO$_2$ for a 12 year period, as an equivalent net-present value (NPV) payment of \$45/tCO$_2$ in 2022 USD over an assumed 30-year asset life in the ``REF'' case. To be consistent with this treatment, we also use a 30-year lifetime for CCS retrofit and new gas turbines. In the ``CCS-20-year lifetime'' case, we instead assume a 20 year asset life for CCS retrofit and new gas turbines, which increases the equivalent value of the credit from \$45 to \$58 per ton of CO\textsubscript{2}. While these values are equivalent in NPV terms, the modeled value of the 45Q credit affects the marginal cost of qualifying generators and thus their position in the economic dispatch. Higher (lower) modeled tax credit values can thus result in more (or less) annual generation (and thus revenue from electricity sales) from a qualifying power plant. We thus observe that modeling the NPV of 45Q over a 20-year asset life increases capacity of coal CCS retrofits and new gas CCS from 1 and 9 GW to 61 and 54 GW, respectively, leading to 30 MtCO$_2$ of emissions reduction in 2040(Figure \ref{fig:capGenEmis_sensitivity}). This finding indicates that CCS technologies are borderline economic for certain existing coal-fired generators and new gas turbines that are eligible for 45Q (and proximate to cost-effective transport and storage). Real-world investment decisions will incorporate more complex assessments of project risk and different agents (with varying risk aversion) may opt for different strategies. 

Across all sensitivities, EPA rules consistently reduce emissions in 2040, by 18-58\% relative to the equivalent ``No Regulations'' benchmark (reaching 34-74\% below 2022 emissions level). Natural gas prices have the most significant impacts on emissions outcomes. With low gas prices, economic coal retirement leads to low emissions even without EPA rule and the implementation of GHG regulations only further reduce 2040 emissions by 18\%. However, when natural gas price is high, more coal-fired generators remain economic and avoid retirement, resulting in substantial emissions in the absence of EPA regulations. We thus observe the greatest emission reductions from 2025 to 2040 in this high gas price scenario, with more than 500 MtCO$_2$ emissions avoided in both 2035 and 2040, relative to no regulations.

\subsection{Alternative strategies to cost-effectively limit emissions from fossil fuel-fired generators may exist.}\label{sec:alternatives}

\begin{figure}[h]
    \centering
    \includegraphics[width=1.0\textwidth]{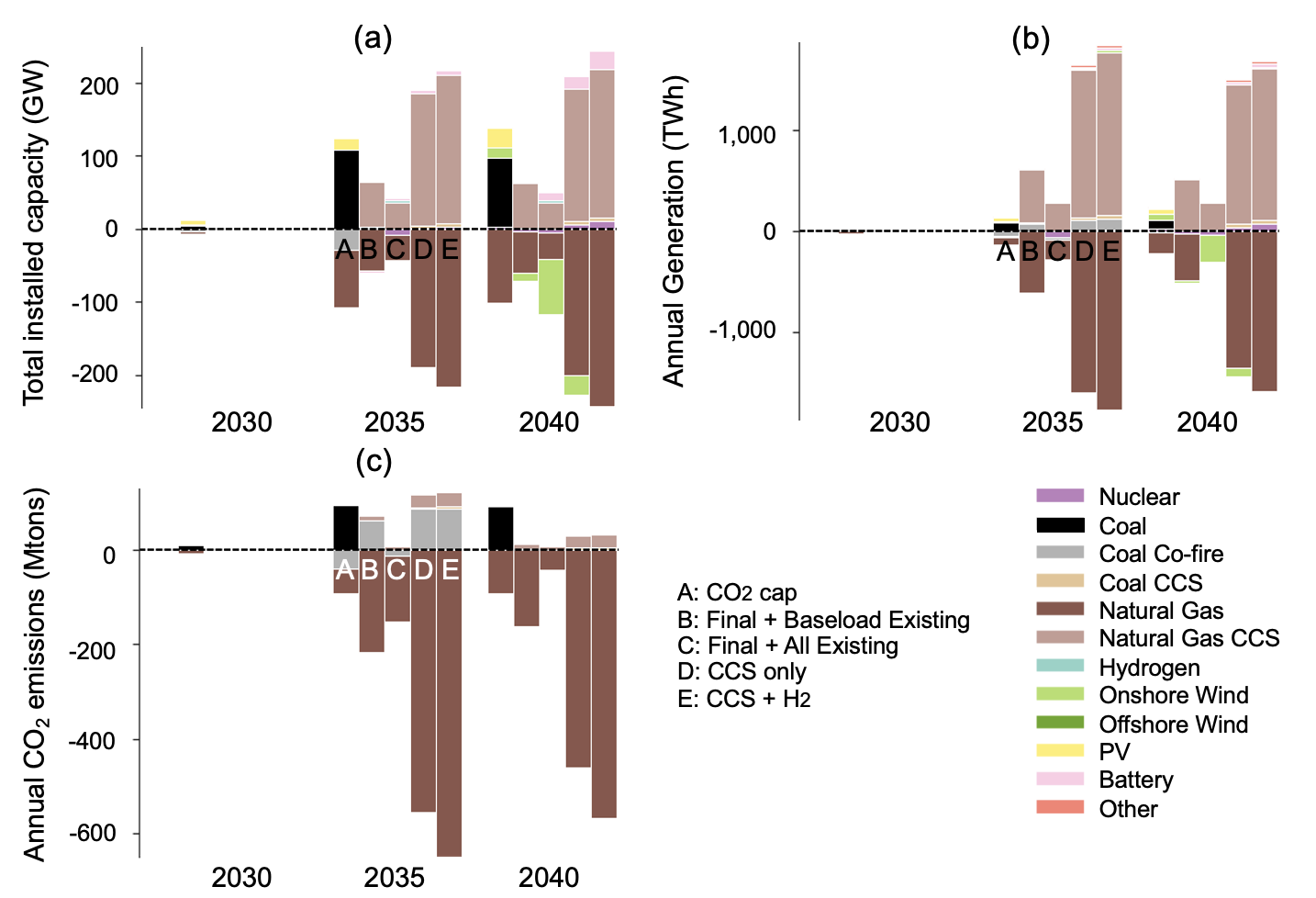}
    \caption{Differences between the ``Coal + New Gas" rule and alternative strategies in total installed capacity (a), electricity generation (b), and CO\textsubscript{2} emissions (c) by technology in each period.}
    \label{fig:capGenEmis_all_alt}
\end{figure}

EPA's GHG regulations require different classes of generators to meet different emissions performance standards, with the performance standard based on several specific compliance options implementable at the plant (i.e., co-firing, carbon capture, etc.). This ``inside-the-fenceline'' approach is intended to be responsive to a 2022 Supreme Court decision\cite{epa_wv} that struck down the more flexible sectoral emissions performance standards proposed by EPA under the Obama Administration (the so-called ``Clean Power Plan''\cite{cpp}). To compare the performance of the current proposed rules to a more flexible sector-wide emissions cap, we also evaluate a new scenario (``CO\textsubscript{2} Cap'') that has a national emission limit equal to total emissions under the final rules (``Coal +  New Gas'') in each period. Compared with the final EPA rules (the ``Coal + New Gas'' scenario), we find that setting an emission cap without any further regulations on the operations of generators results in more generation from existing coal, new natural gas, and renewable resources in lieu of generation from coal plants co-firing with gas or existing gas plants (Figures \ref{fig:capGenEmis_all_alt} and \ref{fig:capGenEmis_NG_alt}). Although the average cost of mitigation is relatively low under the EPA rules (\$17/ton reduced from ``No Regulations'' vs ``Coal + New Gas''), an emissions cap will always exhibit superior static efficiency. Indeed, the average cost to mitigate CO\textsubscript{2} emissions declines 82\% under this emissions cap scenario to just \$3/ton, compared with ``Coal + New Gas'' (Figure \ref{fig:emisByCase}). 

Additionally, since EPA is seeking input and analysis on possible regulations for existing gas-fired generators\cite{epa_delay}, we evaluate several strategies to regulate existing combustion turbines (``Final + Baseload Existing'' \& ``Final + All Existing'' in Table \ref{tab:caseDescr}). In addition to regulations on coal and new gas generators, ``Final + Baseload Existing'' requires CCS retrofit for all existing baseload gas combustion generators ($>$40\% capacity factor), regardless of their nameplate capacity. As a result, overall gas-fired capacity does not significantly change but more new gas with CCS is added to replace existing gas turbines, incrementally reducing 2040 emissions by 152 MtCO$_2$ (reaching 62\% below 2022 emissions) at an average abatement cost of \$12/tCO$_2$, a 29\% improvement in average abatement cost vs the final EPA rule. 

In ``Final + All Existing'', we require \textit{all} existing gas turbines to retrofit with CCS (regardless of capacity factor). This requirement promotes some addition of gas CCS retrofit but most existing gas turbines will still retire and a significant amount of new non-baseload gas turbines that are not subject to CCS requirements are added to the system (Figure \ref{fig:capGenEmis_NG_alt}). Surprisingly, this strict regulation on existing gas-fired generators does not reduce emissions much as emission reductions from more CCS retrofit are offset by emission increases from additional new gas capacity. Building a lot of new gas turbines which cannot run with capacity factor greater than 40\% also roughly doubles the average abatement cost relative to the final EPA rules (``Coal + New Gas''). The incremental cost of reductions achieved relative to ``Coal Only'' amounts to \$87/tCO$_2$ under this strategy (Figure \ref{fig:emisByCase}).

Finally, based on the observation from previous scenarios that imposing strict emission regulations on only one category of gas turbines (new or existing) is likely to increase generation and emissions from the other category, we consider a pair of alternative strategies (``CCS Only'' and ``CCS + H\textsubscript{2}'') that apply equivalent standards to both new and existing turbines. Under ``CCS Only,'' all non-peaker gas-fired generators (capacity factor $>$20\%) must generate electricity with CCS from 2035. Given high costs of H\textsubscript{2}, it is not economic to use H\textsubscript{2} in gas turbines for base load operations, so ``CCS + H\textsubscript{2}" requires co-firing 30\% H\textsubscript{2} in gas peakers ($\le$ 20\% capacity factor) while all other gas turbines have to eventually be equipped with CCS beginning in 2035. In both cases, we keep the constraint on coal-fired generators the same as `` Coal Only.'' Under both rules, most gas turbines retrofit with CCS in 2035 so that they can keep running at a relatively high utilization level. To avoid using expensive H\textsubscript{2}, even more CCS is added under ``CCS + H\textsubscript{2}''. Compared with the EPA rules (''Coal + New Gas''), this conversion to CCS reduces 2040 CO\textsubscript{2} emissions by 430 MtCO$_2$ under ``CCS Only'' and 534 MtCO$_2$ under ``CCS + H\textsubscript{2}'' (reaching 81\% and 88\% below 2022 emissions, respectively). More importantly, extending regulations to all non-peakers (CF $>$20\%) instead of only regulating baseload generators (CF $>$40\%) reduces more emissions than only applying rules to new gas combustion generators (``Coal + New Gas'') or extending current new gas rules to existing gas generators (``Final + Baseload Existing'' and ``Final + All Existing'') at a lower incremental cost of \$16 and \$22 for each ton of reduction above that achieved in the ``Coal Only" case, respectively (Figure \ref{fig:emisByCase}). We do not observe significant increase in renewable capacity addition after applying these strict regulations on natural gas because installed capacity of wind and PV already reaches the maximum permitted in 2035 under ``Coal + New Gas''. More renewable capacity will be added to the system if no renewable growth rate limits are applied. Adding additional renewable capacity principally reduces CCS retrofit capacity, which lowers average abatement costs but does not significantly reduce emissions further (Figure \ref{fig:capGenEmis_noVREmax}).  

\subsection{Discussions}\label{sec4}

In this work, we investigate the potential impacts of EPA power plant emissions performance standards on power system operations and emissions by incorporating the complex operational constraints imposed by the rules for different classes of fossil-fueled generators into a detailed capacity expansion model. We observe significant CO\textsubscript{2} emission reductions from 2035 to 2040 under the finalized rules. Emissions in 2040 reach 51\% below 2022 emissions level under our reference assumptions and 34-74\% under a range of sensitivity analyses conducted on fuel prices, renewable resource availability, and tax credit treatment, demonstrating that the EPA regulations can substantially reduce power sector emissions across a range of plausible future conditions. The most significant reductions occur with high natural gas prices, positioning the EPA rules as an effective backstop against elevated emissions from prolonged operation of unabated coal-fired generators.

By isolating the impacts of the rules applied to different classes of fossil fueled generators, we determine that coal retirement (as a result of regulation on coal-fired steam generators) contributes most to the emission mitigation under EPA's regulations (accounting for 70\% of all reductions from the no regulations benchmark in 2040). Finalized regulations for new gas generators are primarily met by restricting average utilization rate (capacity factor) to $<40\%$ to avoid more costly CCS requirements. Regulations on new gas units further reduce emissions but at a relatively high average cost of abatement: \$75 per ton of additional CO$_2$ reductions vs applying regulations only to existing coal generators. This \textit{de facto} constraint on capacity factors for new, more efficient gas generators results in more overall new gas capacity utilized less frequently, raising the effective average cost of electricity from new gas plants and increasing reliance on inefficient existing gas generators. 

Compared with the initial proposed rules, we find that EPA's decision to delay finalization of regulations on existing large and baseload gas turbines only yields minimal negative emission impacts, as the proposed rule for existing gas only applies to a small share of existing generating capacity, and these units can comply with the rules simply by operating at $<$40\% capacity factor annually. Additionally, given the high price of clean hydrogen, removing the H\textsubscript{2} co-firing requirement for gas generators in the proposed rule reduces average abatement costs. 

To inform EPA's development of future regulations on existing gas generators, we also assess several potential alternative rules to identify opportunities to increase the efficacy or efficiency of emissions regulations. We extend current EPA rules for new gas turbines to cover existing gas generators. Compared with finalized rules, extending equivalent standards for new baseload gas generators to existing baseload gas plants results in a 21\% emission reduction compared with current finalized rules in 2040, with an affordable abatement cost at \$12 per ton of CO\textsubscript{2}. However, we find that requiring all existing gas turbines (regardless of capacity factor) to retrofit with CCS by 2035 does not significantly reduce emissions but further increases the ``overbuild'' of new gas turbines used at low utilization rates to avoid CCS requirements, making electricity generation much more expensive. Under alternative rules requiring (a) all non-peakers (capacity factor $>20\%$) to install CCS or (b) all non-peakers to install CCS \textit{and} all peakers to co-fire 30\% H\textsubscript{2}, we find larger emission reductions at relatively low incremental emission abatement costs compared to  ``Coal Only''. This potential rule would promote the use of CCS on higher utilization rate plants and, by applying consistent standards to new and existing generators, would accelerate the retirement of inefficient generators, resulting in a capacity portfolio closer to that observed under an emission cap, where the power system will achieve emission targets in a most cost-effective way (e.g., using less inefficient NG-fired generators, Figure \ref{fig:capGen_cap}). Compared with current EPA rules, which we find increase the average abatement cost by 470 \% relative to a ``CO\textsubscript{2} Cap'' case achieving an equivalent emissions reduction (\$3 to \$17 per ton of CO\textsubscript{2}), we find that extending CCS requirements to all non-peakers (and additionally H$_2$ co-firing requirements for peakers) only increases average abatement costs by 83\% (or 100\% with the peaker co-firing rule) from \$6 to \$11 (or \$8 to \$16) per ton of CO\textsubscript{2}. These two alternative regulations also achieve the lowest overall emissions in 2040: 81 and 88\% below 2022 emissions level, respectively.

In summary, we find that regulations on gas-fired generators should be carefully considered to avoid overly restricting new, efficient gas generators relative to existing gas-fired units. Applying stricter regulations to only one subcategory of gas turbines (new or existing) will potentially increase gas-fired generation from the other gas subcategory, not necessarily mitigate emissions cost-effectively. The removal of H\textsubscript{2} co-firing requirements on all new gas generators from the proposed rules is a step towards this goal by reducing the utilization of less efficient existing generators. However, the lack of regulations on new non-baseload (CF $\leq$40\%) generators still leads to many generators choosing to lower their utilization level to avoid CCS investment costs. Alternatively, applying consistent emissions regulations to a larger group of gas-fired generators regardless of vintage -- such as requiring all plants with CF $>$20\% to install CCS and all plants with CF $\leq$20\% to co-fire 30\% hydrogen by 2035  -- can level the playing field and increase the emission mitigation efficacy and economic efficiency of the overall regulation (e.g., achieve larger emission reductions and lower average mitigation costs). 
%TC:ignore
\section{Methods}\label{sec3}

\subsection{Power System Modeling}

We analyze the power system of the continental United States (CONUS) with input parameters based on Energy Information Agency (EIA) data (Form EIA-860 (2022)\cite{eia860}, Form EIA-860M (June 2023) \cite{eia860m}, and Form EIA-923 (2022)\cite{eia923}), with natural gas and coal accounting for 45\% and 15\% of total installed capacity, respectively (Figure \ref{fig:CapExisting}). We include around 25,000 electricity generators that exist in the US power grid in 2022 (including both thermal and renewable resources), and to improve computational tractability, we cluster them into $\sim$500 resources groups based on their fuel types, generator types, capacity sizes, heat rates, and locations. Within each resource cluster, generators are treated as identical and share the characteristics of the cluster average. We also include distributed solar in our model\cite{cambium}. To consider transmission constraints, we divide the CONUS into 26 zones based on aggregations of the 64 regions used in the EPA's Integrated Planning Model (IPM) and model 49 transmission paths connecting these zones to form the initial transmission network\cite{ipm}. Each zone is treated as a balancing area (BA) with their own electricity demand and we assume transmission flows are unconstrained within each zone. However, the cost of connecting new wind and solar resources includes estimated costs to interconnect renewable energy project areas to demand centers within each zone using a transmission routing and costing algorithm described in \cite{patankar2023land}. Figure \ref{fig:tx_and_demand} shows the study domain of this work. In all zones, we allow building new generators (nuclear, natural gas turbines with and without carbon capture and storage (NG w/wo CCS), hydrogen-ready turbines (H\textsubscript{2}), onshore and offshore wind, solar PV, and battery storage. We also model retrofitting of existing gas and coal generators with CCS or H\textsubscript{2} co-firing and existing coal plants to co-fire with natural gas. Fuel blend ratios for co-fired plants are optimized within the model (though they are subject to pertinent constraints from EPA regulations in each case). The model also co-optimizes inter-zonal transmission expansion. Inter-zonal transmission expansion costs represent estimated costs to expand high-voltage transmission between a pair of metropolitan statistical areas (MSAs) with $>$1 million population located in neighboring zones (or the largest MSA if none of that size are present), as well as to expand `backbone' transmission networks connecting any MSAs with population $>$1 million within each zone (if more than one is present in the zone)\cite{patankar2023land}. 

To determine what and when to build new generators or retire existing generators, we employ GenX\cite{jenkins2017enhanced, genx}, a state-of-the-art capacity expansion model that considers detailed power plant operational constraints to optimize the system operation and investment decisions in every planning period. We model 52 weeks of operations with hourly resolution to represent each planning period. All input data except for green hydrogen production demand (i.e., operational parameters of existing generators\cite{eia860, eia860m}, renewable energy potential\cite{repeat}, transmission capacity\cite{ipm}, zonal electricity demand\cite{nrel_demand, ferc714}, costs assumptions\cite{nrel_atb2023, aeo2023, eia_short}) are prepared by PowerGenome\cite{powergenome}, an open-source software tool that combines data from Environmental Protection Agency (EPA), National Renewable Energy Laboratory (NREL), and Rapid Energy Policy Evaluation and Analysis Toolkit (REPEAT) to rapidly produce input data for electricity system planning models. Additionally, we include an exogenous regional H\textsubscript{2} production requirement that is met by optimized electrolyzer capacity and operations and assume H\textsubscript{2} is available at gas generators at a specified fuel price. H\textsubscript{2} has to be produced by electrolysis powered by new renewable energy matched with supply at an hourly basis within the same model zone, as per proposed Treasury Department guidance for the 45V hydrogen tax credit\cite{45v}. We obtain the regional H\textsubscript{2} demand and investment cost assumptions for electrolyzers from Haley et al. (2023)\cite{adp2023}. We model key incentives provided by the Inflation Reduction Act of 2022 (IRA), particularly the production and investment tax credits for new carbon-free electricity, production tax credit for existing nuclear (modeled as preventing economic retirement through expiration in 2032), and the 45Q tax credit for carbon capture and storage\cite{bistline2023emissions}. Electricity demand profiles and levels (from \cite{repeat}) also account for impacts of IRA incentives on electrification of transportation and space and water heating. For model validation, we run GenX with installed capacities and historical fuel prices from 2021 without any capacity expansion and determine EGU operations and system emissions. Electricity generation shares by resource type are very close to historical data from EIA (Figure \ref{fig:2021}), indicating that GenX reproduces realistic market dynamics, including capturing economic competition between coal and gas generators, and is thus suitable for this kind of analysis.

As the rules proposed by EPA regulate emissions from fossil-fuel fired power plants from now to 2038, we investigate their impacts on the power system until 2040, modeling four periods 2023 - 2025 (2025), 2026 - 2030 (2030), 2031 - 2035 (2035), and 2036 - 2040 (2040). Using capacity outputs from the the previous period as existing capacity in the system, the capacity expansion in each period is optimized sequentially (``myopic'' mode). Unlike ``perfect foresight``, which makes decisions across the entire temporal domain at high computational costs, ``myopic`` mode minimizes annuitized capital investments and variable costs in every period, but does not anticipate future periods when planning each stage. For a comparison with the existing power system in 2022, we use installed capacity and electricity generation of each EGU from Form EIA-923 and calculate system emissions with constant emission rates for coal and natural gas, respectively.

\subsection{Modeling EPA rules}
EPA allows generators to choose different compliance pathways to reduce emissions (i.e., CCS retrofit, reduce utilization level, or H\textsubscript{2} co-firing (in proposed rules)) but those mitigation requirements vary by operational status and generator types. For example, EPA's performance standard for existing coal-fired generators that would like to operate after 2039 is based on retrofitting to equip CCS from 2032, but the standard for coal-fired generators that would retire before 2039 is based on co-firing 40\% natural gas (in heat input) starting from 2030. Therefore, we run a benchmark case with GenX to determine the economic retirement period of all coal generators absent EPA regulations and apply different constraints to each coal unit based on this economic retirement schedule. The finalized EPA rules can be grouped into two categories (``Coal'' and ``New Gas'') applying to existing coal-fired boilers and new combustion turbines, respectively. Figure \ref{fig:EPA_rules} shows how those regulations are modeled in GenX. Regulations for existing natural gas-fired combustion turbines included in the proposed rules, the regulations will vary by both nameplate capacity and capacity factor.

To consider detailed regulations on different types of generators, we enhance GenX by allowing generators to burn multiple fuels in one turbine and introducing more operational constraints (e.g., minimum co-fire level and maximum capacity factor) to represent conditions set by the EPA regulations. See Table \ref{tab:equations} for details.

To calculate the average CO\textsubscript{2} mitigation cost under each scenario, we compare increases in annualized system costs and reductions in CO\textsubscript{2} emissions reductions relative to the ``No Regulations'' case. We also calculate the average cost of incremental reductions relative to the ``Coal only'' scenario. We also include sensitivity analyses on renewable growth rates, natural gas prices, coal prices, and tax credit treatment, to understand how future uncertainties would affect the emission reduction potential of the proposed rules (Figures \ref{fig:VREgrowth} and \ref{fig:fuel_prices}).

\subsection{Alternative strategies}

In addition to rules proposed by EPA, we include three groups of alternative mitigation strategies that have an emission limit similar to modeled emissions from including all EPA rules (``CO\textsubscript{2} Cap''), extend finalized EPA rules to existing gas-fired generators in the system (``Final + Baseload Existing'' \& ``Final + All Existing''), or explore alternative regulations for all gas turbines (``CCS Only'' and ``CCS + H\textsubscript{2}''). Table \ref{tab:caseDescr} provides a detailed description of these alternatives. All alternatives except for ``CO\textsubscript{2} Cap''  apply the same regulations on coal as in the final EPA rules.

\section{Acknowledgement}

Funding for this work was provided by the Princeton Zero-carbon Technology Consortium, supported by unrestricted gifts from GE, Google, ClearPath, and Breakthrough Energy.

%TC:endignore
\newpage

\bibliography{sn-bibliography}% common bib file

\end{document}